\documentclass[english]{bvm2019}

\title{Deep Learning Algorithms for Coronary Artery Plaque Characterisation from CCTA Scans}
\titlerunning{Deep Learning for Plaque Characterisation}

\author{Felix Denzinger$^{1,2}$, Michael Wels$^2$, Katharina Breininger$^1$, Anika Reidelsh\"ofer$^3$, Joachim Eckert$^4$, Michael S\"uhling$^2$, Axel	Schmermund$^4$, Andreas Maier$^1$}

\authorrunning{Denzinger et al.}

\institute{%
$^1$Pattern Recognition Lab, Friedrich-Alexander-University Erlangen-Nuremberg,
Erlangen, Germany\\
$^2$Computed Tomography, Siemens Healthineers, Forchheim, Germany\\
$^3$University Clinic Frankfurt, Frankfurt am Main, Germany\\
$^4$Cardioangiological Centrum Bethanien, Frankfurt am Main, Germany\\}

\email{felix.denzinger@fau.de}

\usepackage{ulem}
\usepackage{xcolor}
\usepackage{xspace}

\newcommand{\tble}[2]{#1\scriptsize{\ensuremath{\pm}#2}}
\newcommand{\tblee}[2]{\textbf{#1\scriptsize{\ensuremath{\pm}#2}}}

\begin{document}

%
\selectlanguage{english}

\maketitle

\begin{abstract}
	Analysing coronary artery plaque segments with respect to their functional significance and therefore their influence to patient management in a non-invasive setup is an important subject of current research. 
	In this work we compare and improve three deep learning algorithms for this task: A 3D recurrent convolutional neural network (RCNN), a 2D multi-view ensemble approach based on texture analysis, and a newly proposed 2.5D approach. 
	Current state of the art methods utilising fluid dynamics based fractional flow reserve (FFR) simulation reach an AUC of up to 0.93 for the task of predicting an abnormal invasive FFR value. For the comparable task of predicting revascularisation decision, we are able to improve the performance in terms of AUC of both existing approaches with the proposed modifications, specifically from 0.80 to 0.90 for the 3D-RCNN, and from 0.85 to 0.90 for the multi-view texture-based ensemble. The newly proposed 2.5D approach achieves comparable results with an AUC of 0.90.
	
\end{abstract}

\section{Introduction}
Cardiovascular diseases (CVDs) remain the leading cause of natural death~\cite{mendis15}. In diagnosis and treatment of CVDs, the identification of functionally significant atherosclerotic plaques that narrow the coronary vessels and cause malperfusion of the heart muscle plays an important role. In clinical practice, this is typically assessed using fractional flow reserve (FFR) measurements~\cite{Cury16}.

This measurement is performed minimally invasively and therefore induces a small but existing risk to the patient. A non-invasive modality capable of visualising and assessing coronary artery plaque segments is coronary computed tomography angiography (CCTA). Current research tries to simulate the FFR value from CCTA scans \cite{taylor13}. Approaches based on this mostly rely on a prior segmentation of the whole coronary tree which is computationally intensive, prone to errors and may need manual corrections \cite{wels16}. 

In this work, we investigate three lumen-extraction independent deep learning algorithms for the task of predicting the revascularisation decision and the significance of a stenosis on a lesion-level. We propose a multi-view 2.5D approach, which we compare with two previously published methods, a 3D-RCNN approach~\cite{zreik18b} and a multi-view texture-based ensemble approach ~\cite{tejero19}. Additionally, we introduce adaptions to improve the performance of all approaches on our task.
These include resizing lesions to an intermediate length instead of padding them and the usage of test-time augmentations.
Also, we propose to use a different feature extraction backbone than described in \cite{tejero19} for the respective approach. 
Note that both reference approaches were originally used to detect lesions and characterise them. Contrary to this we characterise annotated lesions with a defined start and end point.

\section{Material and Methods}
\subsection{Data}
The data collection used contains CCTA scans from 95 patients with suspected coronary artery disease taken within 2 years at the same clinical site. 
For each patient, the resulting clinical decision regarding revascularisation was made by trained cardiologists, based on different clinical indications. This decision was monitored on a branch level. 
Lesions were annotated using their start and end point on the centerline, which was extracted automatically using the method described in \cite{zheng13}.
We binarise the stenosis grade, which is estimated based on the lumen segmentation and defined as the ratio between the actual lumen and an estimated healthy lumen, using a threshold of 50$\,$\% according to \cite{Cury16}.
The branch-wise revascularisation decision is propagated only to the lesion with the highest stenosis grade in branches known to be revascularised. 
Of the total of 345 lesions in our data set, 85 lesions exhibit a significant stenosis grade, and 93 require revascularisation.

\subsection{Methods}

\subsubsection{3D-RCNN}

The first network we use is identical to the method described in \cite{zreik18b}. 
In this approach, after extracting the coronary centerlines, a multi-planar reformatted (MPR) image stack is created by interpolating an orthogonal plane for each centerline point. Next, the MPR image stack is cut into a sequence of 25 overlapping cubes with size 25x25x25 and a stride of 5. During training, data augmentation using random rotations around the centerline and random shifts in all directions is used. Moreover, the data set is resampled for batch creation to achieve class balance during training. Since detection instead of sole characterisation is performed in \cite{zreik18b}, padding the inputs to the same length was not needed in their work. 

\subsubsection{Texture-based Multi-view 2D-CNN}

The second baseline approach is described in reference \cite{tejero19}. A VGG-M network backbone pretrained on the ImageNet challenge dataset is used as a texture-based feature extractor. The extracted features are encoded as Fisher vectors and used for classification using a linear support vector machine. As inputs for this classification pipeline, different 2D views of the MPR image stack are combined for a final vote.

\subsubsection{2.5D-CNN}
\begin{figure}
	\centering
	\includegraphics[width=0.99\textwidth]{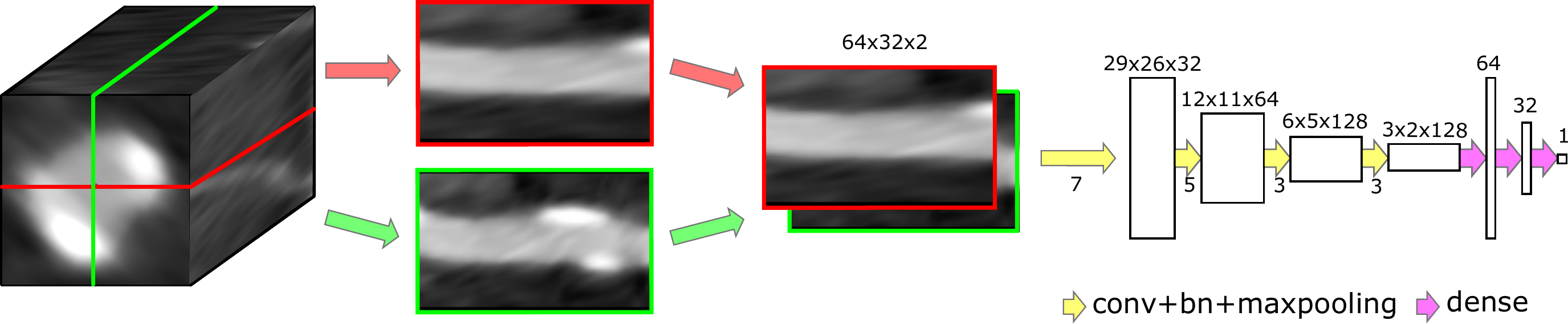}
	\caption{Algorithm overview: Extraction of two orthogonal views of the lesion of interest. These are concatenated and then used as an input for a 2D-CNN (conv = convolutional layer, bn = batch normalisation layer, dense = fully connected layer).} \label{fig1}
\end{figure}

Both aforementioned methods utilize a sliced 3D representation of the lesion or a multitude of 2D representations, which is computationally expensive to obtain and to process by the subsequent machine learning pipeline. To mitigate this, we propose a 2.5D multi-view approach as shown in Figure~\ref{fig1}.
From the MPR image stack, only two orthogonal slices are selected, concatenated and forwarded to a 2D-CNN. 

\subsubsection{Modifications}
In this work, we examine the effect of three different padding strategies for all three approaches: zero-padding, stretching the volume stack to the longest lesion and resizing all lesions to an intermediate size. 
Stretching and squeezing of the image stacks along the centerline is performed with linear interpolation.
Each MPR image stack for each lesion has a resolution of 64x32x32 and 170x32x32 after padding depending on the method used. For the 3D-RCNN approach, we downscale the y and x dimension further to 25x25 to match the original algorithm described in \cite{zreik18b}.
For augmentation of the data set all single volumes are rotated around the centerline in steps of 20$^\circ$, which leads to an 18 times larger data collection. 
In order to create valid rotational augmentations of the image stack without cropping artefacts, we cut out a cylindrical ROI and set all values around it to zero. We confirmed in preliminary experiments that this computationally cheaper procedure does not to impact the results compared to cutting out a rotated view from the original data. In contrast to \cite{zreik18b,tejero19}, no class resampling was necessary during training, since the class imbalance is not as severe for classification given the start and end point of a lesion compared to detecting lesions as well. Instead of the originally proposed VGG-M backbone used in~\cite{tejero19}, we use the VGG-16 network architecture as a backbone since it was already shown to yield better performance in the original paper on texture-based filter banks~\cite{fisher15}. The data set was normalised to fit ImageNet statistics. We also evaluate the performance of this approach using a pretrained Resnet50 architecture \cite{resnet16} as backbone.


\subsubsection{Evaluation}
No hyperparameter optimisation is performed. Parameters are either taken from the references or default values are used.
To reduce the influence of random weight initialisation and other random effects on the results, we repeat a 5-fold cross validation with five different initialisations, leaving a total of 25 splits. All splits are performed patient-wise. We also use the aforementioned rotational augmentation during test-time, and compare how the mean prediction over all rotations performs in comparison to a single input.

\section{Results}

\begin{figure}
	\centering
	\includegraphics[width=0.49\textwidth]{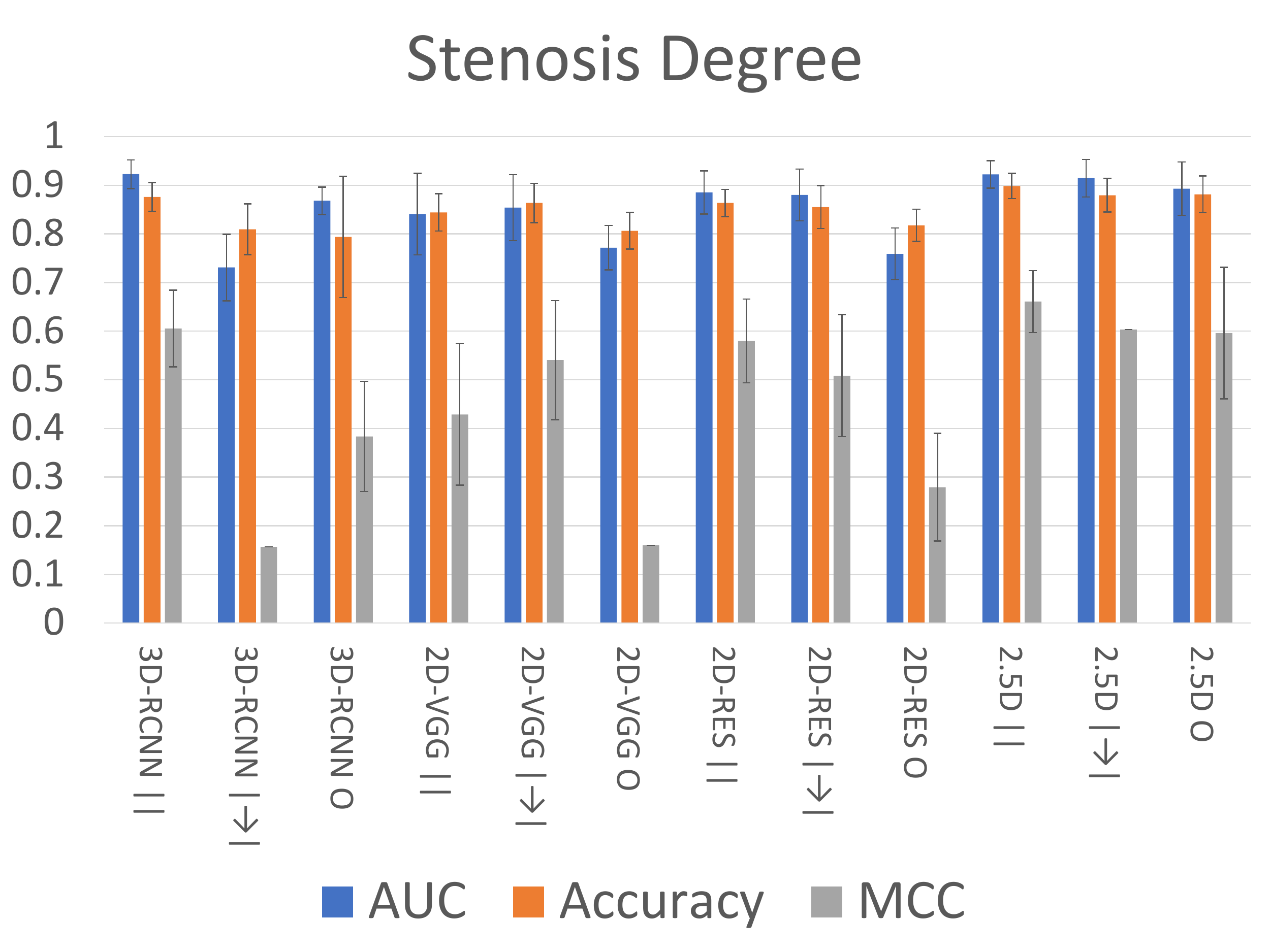}
	\includegraphics[width=0.49\textwidth]{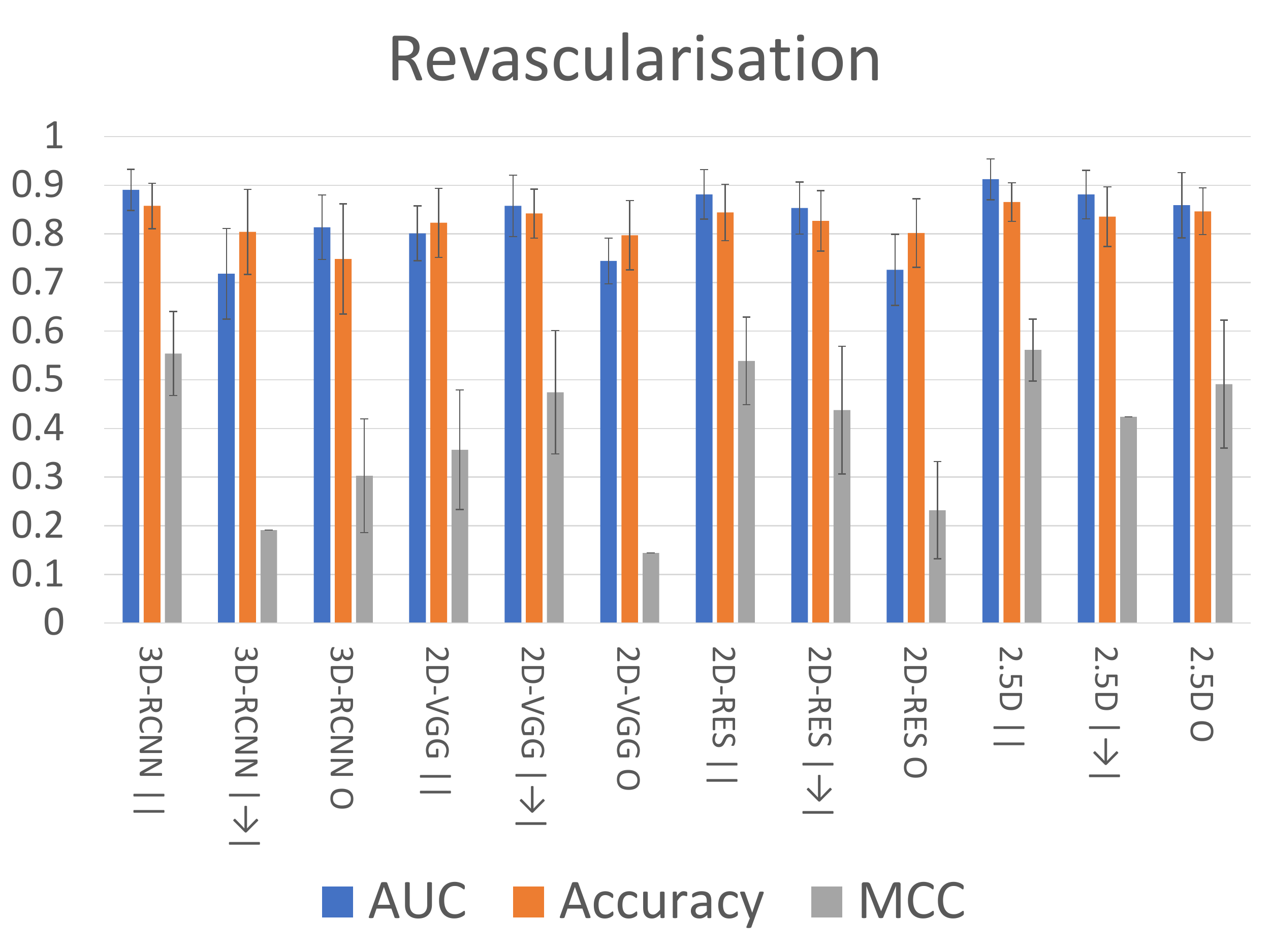}
	\caption{Mean performance and standard deviation of all approaches for different padding strategies. These experiments are performed using only 8 instead of the 18 views (${\vert~~\vert}$ = resizing to intermediate size, ${\vert\to\vert}$ = resizing to the longest sequence, O = zero padding or no padding for the texture-based approach, MCC = Matthews correlation coefficient).} \label{padding}
\end{figure}
\begin{table}
	\centering
	\caption{Results for predicting stenosis degree prediction on a lesion-level ($18$ and $8$ correspond to the amount of views considered for data augmentation during training and test time, $+$ = single view classification, $\ast$ = combined view classification, ${\vert~~\vert}$ = resizing to intermediate size, ${\vert\to\vert}$ = resizing to the longest sequence).}
	\label{stenosisResults}
	\begin{tabular}{ l | l | l | l | l | l | l }
		\hline
		Model/Metric &  AUC & Accuracy & F1-score & Sensitivity & Specificity & MCC \\
		\hline
		\hline
		3D-RCNN\cite{zreik18b}\cite{denzinger19}& 0.89 &	0.85 &  0.67 &	\textbf{0.79}&	0.86&	0.59 \\ 
		
		\hline
		3D-RCNN\cite{zreik18b}\footnotemark[18]$^{\vert~~\vert}$& \tblee{0.92}{0.03} &	\tble{0.88}{0.02} &  \tble{0.69}{0.06} &	\tble{0.68}{0.11}&	\tble{0.93}{0.03}&	\tble{0.62}{0.06} \\ 
		\hline
		2D\cite{tejero19}\footnotemark[8]$^\ast$$^{\vert\to\vert}$$^{VGG}$& \tble{0.85}{0.07} &	\tble{0.86}{0.04} &  \tble{0.62}{0.10} &	\tble{0.56}{0.15}&	\tble{0.94}{0.02}&	\tble{0.54}{0.12}\\	
		\hline
		2D\cite{tejero19}\footnotemark[18]$^{+}$$^{\vert~~\vert}$$^{RES}$ & \tble{0.78}{0.04} &	\tble{0.82}{0.03} &  \tble{0.61}{0.05} &	\tble{0.70}{0.08}&	\tble{0.85}{0.03}&	\tble{0.50}{0.06}\\	
		\hline
		2D\cite{tejero19}\footnotemark[18]$^\ast$$^{\vert~~\vert}$$^{RES}$& \tble{0.90}{0.04} &	\tble{0.87}{0.03} &  \tble{0.68}{0.08} &	\tble{0.71}{0.13}&	\tble{0.91}{0.03}&	\tble{0.60}{0.09}\\	
		\hline
		2.5D\footnotemark[18]$^{+}$$^{\vert~~\vert}$ & \tblee{0.92}{0.03} &	\tble{0.89}{0.02} &  \tble{0.70}{0.06} &	\tble{0.64}{0.10}&	\tble{0.95}{0.03}&	\tble{0.64}{0.06}\\		
		\hline
		2.5D\footnotemark[18]$^\ast$$^{\vert~~\vert}$ & \tblee{0.92}{0.03} &	\tblee{0.90}{0.02} &  \tblee{0.71}{0.07} &	\tble{0.64}{0.10}&	\tblee{0.96}{0.03}&	\tblee{0.66}{0.08}\\			
		\hline
	\end{tabular}
\end{table}
\begin{table}[t]
\begin{minipage}{\textwidth}
	\centering
	\caption{Results for predicting the revascularisation decision on a lesion-level (Abbreviations as in Table~\ref{stenosisResults}).}\label{revascResults}
	\begin{tabular}{ l | l | l | l | l | l | l }
		\hline
		Model/Metric & AUC & Accuracy & F1-score & Sensitivity & Specificity & MCC\\
		\hline
		\hline
		3D-RCNN\cite{zreik18b}\cite{denzinger19}& 0.80 &	0.76 &  0.55 &	\textbf{0.72}&	0.77&	0.42 \\ 
		\hline
		3D-RCNN\cite{zreik18b}\footnotemark[18]$^{\vert~~\vert}$ & \tble{0.90}{0.05} &	\tble{0.84}{0.10} &  \tble{0.63}{0.10} &	\tble{0.65}{0.13}&	\tble{0.90}{0.12}&	\tble{0.53}{0.11} \\ 
		\hline
		2D\cite{tejero19}\footnotemark[8]$^\ast$$^{\vert\to\vert}$$^{VGG}$& \tble{0.86}{0.06} &	\tble{0.84}{0.05} &  \tble{0.56}{0.12} &	\tble{0.49}{0.16}&	\tble{0.93}{0.02}&	\tble{0.47}{0.13}\\	
		\hline
		2D\cite{tejero19}\footnotemark[18]$^{+}$$^{\vert~~\vert}$$^{RES}$ & \tble{0.77}{0.06} &	\tble{0.81}{0.03} &  \tble{0.60}{0.06} &	\tble{0.68}{0.12}&	\tble{0.84}{0.02}&	\tble{0.48}{0.07}\\	
		\hline
		2D\cite{tejero19}\footnotemark[18]$^\ast$$^{\vert~~\vert}$$^{RES}$ & \tble{0.90}{0.06} &	\tble{0.85}{0.05} &\tble{0.66}{0.07} &	\tble{0.70}{0.16}&	\tble{0.89}{0.04}&	\tble{0.57}{0.10}\\	
		\hline
		2.5D\footnotemark[18]$^{+}$$^{\vert~~\vert}$& \tblee{0.90}{0.04} &	\tble{0.87}{0.05} &  \tble{0.65}{0.05} &	\tble{0.60}{0.11}&	\tble{0.94}{0.04}&	\tble{0.58}{0.06}\\	
		\hline
		2.5D\footnotemark[18]$^\ast$$^{\vert~~\vert}$& \tblee{0.90}{0.04} &	\tblee{0.88}{0.05} &  \tblee{0.67}{0.06} &	\tble{0.61}{0.11}&	\tblee{0.95}{0.04}&	\tblee{0.60}{0.07}\\
		
		\hline
	\end{tabular}
\end{minipage}

\end{table}
The most important results are provided in Table~\ref{stenosisResults}, Table~\ref{revascResults} and Figure~\ref{padding}. The results for the 3D-RCNN approach are also compared to the results of our previous work \cite{denzinger19}, where similar experiments are performed on the same data set as here but with the workflow described in \cite{zreik18b}, zero-padding and a different cross validation strategy.
From the three padding methods examined, resizing all volume stacks of the data collection to one intermediate size yields the best results for most network approaches except for the texture-based approach with the VGG-16 backbone, where resizing all lesions to the size of the largest volume performs best. Interestingly, the same algorithm workflow with the Resnet50 backbone performs differently in that regard. A hypothesis that can be drawn from the intermediate padding performing best is that this scale provides on the one hand roughly the same amount of information per sample while on the other hand also keeping the input size in a range where it can be processed better. 
For the 3D-RCNN, we only look at classification in this work, in contrast to the task in \cite{zreik18b} which included the detection of lesions. For this target, the proposed adaptations to the workflow in terms of padding strategy and not resampling the data set during batch creation improves the performance of both predicting the stenosis degree and the revascularisation decision from an AUC of 0.89 to 0.92, and 0.80 to 0.90, respectively.
Having a more powerful feature extractor network for the texture-based approach combined with slightly more data augmentation improves the AUC by 0.05 for classifying stenosis significance, and by 0.04 for classifying revascularisation decision. The method performs considerably better when using test augmentations than without.
Our proposed approach performs similar to the other two approaches, outperforming them by a small margin with an AUC of 0.92/0.90 for predicting a significant stenosis/revascularisation decision. Interestingly, test augmentations only yield a small improvement. This suggests that the method already has all necessary information to predict the task at hand from two orthogonal slices.

\section{Discussion}
In this paper, we compared and improved three segmentation independent deep learning-based algorithms for predicting both significant stenosis degree and clinical revascularisation decision for lesions annotated with a start and end point. We obtained comparable results for each method. Our proposed method -- a 2.5D approach -- slightly outperforms the other approaches and requires fewer views compared to the method previously described in \cite{tejero19}. Therefore, a faster training procedure and inference is possible. In future work, we will examine whether this method is also capable of detecting lesions instead of just classifying them, and whether it is able to predict an abnormal FFR value.

\subsubsection*{Disclaimer}
The methods and information here are based on research and are not commercially available.

\bibliographystyle{bvm2019}

\bibliography{0000}
\end{document}